\UseRawInputEncoding
\documentclass[%
 reprint,
superscriptaddress,
 amsmath,amssymb,
 aps,
prb,
]{revtex4-1}

\usepackage{graphicx}
\usepackage{dcolumn}
\usepackage{bm}
\usepackage{amsmath,amssymb}
\usepackage{graphicx}
\usepackage{color}

\usepackage{hyperref}

\begin{document}

\preprint{AIP/123-QED}

\title{On-demand Plasmon Nanoparticle-Embedded Laser-Induced Periodic Surface Structures (LIPSSs) on Silicon for Optical Nanosensing}

\author{Yulia Borodaenko}
\affiliation{Institute of Automation and Control Processes, Far Eastern Branch, Russian Academy of Science, 5 Radio Str., Vladivostok 690041, Russia}

\author{Sergey Syubaev}
\affiliation{Institute of Automation and Control Processes, Far Eastern Branch, Russian Academy of Science, 5 Radio Str., Vladivostok 690041, Russia}
\affiliation{Far Eastern Federal University, Vladivostok, Russia}

\author{Evgeniia Khairullina}
\affiliation{Saint Petersburg State University, 7/9 Universitetskaya nab., St. Petersburg 199034, Russia}

\author{Ilya Tumkin}
\affiliation{Saint Petersburg State University, 7/9 Universitetskaya nab., St. Petersburg 199034, Russia}

\author{Stanislav Gurbatov}
\affiliation{Institute of Automation and Control Processes, Far Eastern Branch, Russian Academy of Science, 5 Radio Str., Vladivostok 690041, Russia}
\affiliation{Far Eastern Federal University, Vladivostok, Russia}

\author{Aleksandr Mironenko}
\affiliation{Institute of Chemistry FEB RAS, 159 Pr. 100-let Vladivostoka, Vladivostok 690022, Russia}

\author{Eugeny Mitsai}
\affiliation{Institute of Automation and Control Processes, Far Eastern Branch, Russian Academy of Science, 5 Radio Str., Vladivostok 690041, Russia}

\author{Evgeny Modin}
\affiliation{CIC NanoGUNE BRTA, Avda Tolosa 76, 20018 Donostia-San Sebastian, Spain}

\author{Evgeny L. Gurevich}
\email{gurevich@fh-muenster.de}
\affiliation{Laser Center (LFM), University of Applied Sciences M\"unster, Stegerwaldstra?e 39, 48565 Steinfurt, Germany}

\author{Aleksandr A. Kuchmizhak}
\email{alex.iacp.dvo@mail.ru}
\affiliation{Institute of Automation and Control Processes, Far Eastern Branch, Russian Academy of Science, 5 Radio Str., Vladivostok 690041, Russia}
\affiliation{Pacific Quantum Center, Far Eastern Federal University, Vladivostok, Russia}

\begin{abstract}
 Ultrashort laser pulses allows to deliver electromagnetic energy to matter causing its localized heating that can be used for both material removal via ablation/evaporation and drive interface chemical reactions. Here, we showed that both mentioned processes can be simultaneously combined within straightforward laser nanotexturing of Si wafer in functionalizing solution to produce a practically relevant metal-semiconductor surface nano-morphology. Such unique hybrid morphology represent deep-subwavelength Si LIPSSs with a extremely short period down to 70 nm with their high-aspect-ratio nanotrenches loaded with controllable amount of plasmonic nanoparticles formed $via$ laser-induced decomposition of the precursor noble-metal salts. Moreover, heat localization driving reduction process was utilized to produce surface morphology locally decorated with dissimilar plasmon-active nanoparticles. Light-absorbing deep-subwavelength Si LIPSSs loaded with controllable amount of noble-metal nanoparticles represent an attractive architecture for plasmon-related applications such as optical nanosensing where efficient coupling of the propagating optical waves to highly localized electromagnetic ``hot spots`` is a mandatory requirement. To support this statement we demonstrated applicability of such hybrid morphology for fluorescence-based detection of nanomolar concentrations of mercury cations in solution.
\end{abstract}

\maketitle

\section{Introduction}

Nanostructures made of noble-metals are widely used to enhanced interaction of the optical radiation with a matter giving rise to plethora of practically relevant interface phenomena such as surface-enhanced Raman scattering (SERS), emission, higher harmonic generation and catalysis \cite{aslan2005metal,balvcytis2018fundamental,langer2019present,wu2019plasmonic,linic2011plasmonic}. The mentioned effects are largely associated with enhanced absorption of the electromagnetic (EM) energy by such nanostructures that can support coherent oscillations of free electron plasma, i.e. surface plasmons \cite{maier2007plasmonics}. Such hybrid EM waves can be localized in the close proximity to the surface of the noble-metal nanostructures as well as propagate along the metal-dielectric interfaces. In both cases, local amplitude of the plasmon-mediated EM field can be substantially enhanced, that is highly demanded for various abovementioned applications. However, the noble metals suffer from high dissipative losses, excessive light-to-heat conversion efficiency and incompatibility with common lithography-based fabrication technologies limiting the performance of plasmon-related approaches and practical realization of the real-life devices. Beyond application of plasmon-active nanostructures, metal-free nanophotonic platforms were recently suggested to empower light-matter interaction at the nanoscale \cite{kuznetsov2016optically}. Such all-dielectric concept is based on the light localization by nanostructures made of dielectric and semiconductor materials characterized by rather low optical losses and high refractive index such as Si, Ge, TiO$_2$. Resonant absorption, excitation of Mie resonances as well as localization of light in the gaps between such high-index nanostructures were proven to be useful for various photonic, optoelectronic and sensing applications \cite{alessandri2016enhanced,mitsai2018chemically,mironenko2019ultratrace,koshelev2020subwavelength,yesilkoy2019ultrasensitive,chen2020flat,leitis2021wafer}.

Meanwhile, owing to the lack of materials with desired optical characteristics, the integration of the plasmonic and all-dielectric paradigms within practically relevant unified (hybrid) nanostructures is considered as a next step toward realization of advanced devices with expanded functionality \cite{jiang2014metal}. From this point of view, combination of basic plasmonic materials (such as Au or Ag) with silicon (Si) that ideally fit the requirements of optical-range nanophotonics, is extremely promising for light-matter applications. However, practical realization of the hybrid nanostructures and related devices is hindered by the lack of cost-effective single-step and maskless fabrication approaches for integration of such dissimilar materials. Apart from lithography-based techniques irreplaceable in microelectronics, direct laser processing employing interaction of pulsed laser radiation with a matter has recently appeared as a straightforward application-ready solution for nanofabrication of various functional nanostructures and nanotextured surfaces (including those made of such dissimilar metal-semiconductor materials) \cite{sugioka2014ultrafast,malinauskas2016ultrafast}.

Utilization of femtosecond (fs) pulses allows to precisely deliver laser energy into the material at minimized heat affect zone, while ultrafast lattice thermalization ensures localized phase transitions and delicate material removal by ablation or evaporation. Multi-pulse material exposure permits to observe self-organization phenomena driven either by excitation of EM waves at the interface of the photoexcited material and surroundings or hydrodynamic instabilities that result in formation of practically relevant grating-type surface morphologies, referred to as laser-induced periodic surface structures (LIPSSs) \cite{buividas2014surface,bonse2016laser,bonse2020maxwell}. Processing conditions (such as laser fluence, pulse duration and repetition rate, ambient surroundings, etc.) permit to control the LIPSS morphology and periodicity especially important for applications, while the grating orientation is typically defined by polarization vector of the incident laser radiation.

Interestingly, since the first observation of LIPSS dated back to 1965 \cite{birnbaum1965semiconductor}, Si was among the most studied materials regarding LIPSS formation. Recent research efforts achieved regular Si morphologies with near-wavelength periodicity (low-spatial frequency LIPSSs) \cite{huang2019fabrication,dostovalov2020hierarchical,bronnikov2021uniform} or shrunk the periodicity down to deep-subwavelength values (high-spatial frequency LIPSSs) of $\approx\lambda$/10 ($\lambda$ is a laser radiation wavelength) \cite{miyaji_mechanism_2012,le2011generation,le2005sub,zhang2019hierarchical,hamad2014femtosecond,shen2008high,yiannakou2017cell,kesaev2021nanopatterned,straub2012periodic,zhang2019hierarchical,borodaenko2021deep}. Several studies were directed to produce hybrid metal-semiconductor morphologies by LIPSS patterning of bi-layer samples \cite{reinhardt2015directed,reinhardt2015highly}, yet without discussion of the optical properties and related applications. Two-step approaches were also suggested: deep-subwavelength Si LIPSSs created in the first step were later loaded with chemically sensitized plasmonic nanoparticles (NPs) \cite{hamad2018femtosecond} or over-coated with nanometer-thick metal films \cite{diebold2009femtosecond,borodaenko2021deep} to create SERS-active substrate for molecular detection. Finally, to the best of our knowledge, only a few studies reported straightforward fabrication of Si LIPSS decorated by either Au \cite{broadhead2020fabrication,li2020shaped} or Ag \cite{lin2009one} NPs by fs-laser radiation simultaneously driving Si texturing in liquid environment as well as nanoparticle synthesis via plasma-assisted reduction process.

Here, adopting similar approach we demonstrated precise and highly controllable deep-subwavelength LIPSS nanotexturing of Si wafer in distilled water containing either AgNO$_3$ or HAuCl$_4$. Delicate below-threshold fs-laser texturing allowed to produce plasmon-nanoparticle-embedded Si LIPSS with unique morphology (the shortest ever reported period of 70 nm and depth-to-period ratio as high as 3) revealed by combining transmission electron microscopy (TEM), energy-dispersive X-ray (EDX) characterization and tomographic reconstruction. The LIPSS formation also localizes laser-induced reduction of both cations to metallic phase allowing to fabricate hybrid metal-semiconductor surface morphologies locally decorated with dissimilar noble-metal NPs. Deep-subwavelength Si LIPSSs loaded with nanoparticles produced by single-step scalable approach is an attractive platform for plasmon-related applications like optical nanosensing which requires efficient coupling of the propagating optical waves to highly localized EM ``hot spots''. By loading these hot spots with an ionochromic light-emitting molecule, Rhodamine 6G thiohydrazide, we realized an optical sensor for detection of trace concentrations (0.15 nM - 100 nM) of Hg$^{2+}$ ions in water based on surface-enhanced fluorescence (SEF) effect.

\begin{figure}[t!]
\centering
\includegraphics[width=1.\columnwidth]{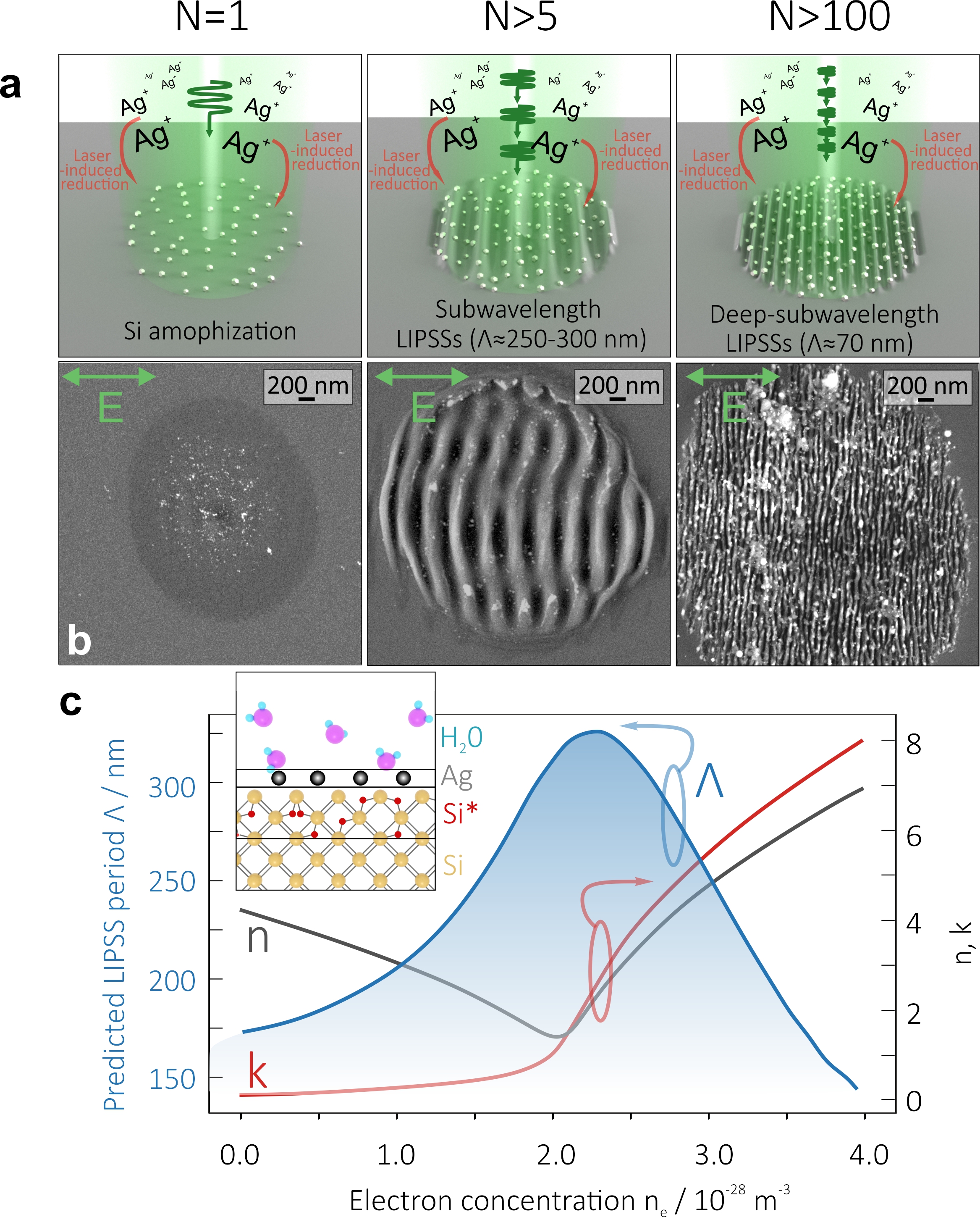}
\caption{(a) Schematic representation of Si surface evolution upon single- and multi-pulse exposure in aqueous solution of AgNO$_3$ (10$^{-3}$ M) assisted by simultaneous formation of the Ag NPs $via$ laser-induced reduction. (b) Top-view SEM images of the Si surface irradiated at fixed fluence $F$=0.1 J/cm$^2$ by varying the number of laser pulses per spot $N$: $N$=1 (left), 10 (middle) and 1000 (right). (c) LIPSS period $\Lambda$ (blue curve), refractive index $n$ (gray curve) and extinction coefficient $\kappa$ (red curve) of the photoexcited Si in water $versus$ the electron concentration n$_{e}$.
}
\label{fig:1}
\end{figure}

\section{Results and discussions}
\subsection{Formation and characterization of nanoparticle-embedded Si LIPSSs}

Formation procedure of the nanoparticle-decorated LIPSSs on the Si surface is schematically illustrated in Figure 1a. The Si wafer is placed in the distilled water containing silver nitrate (AgNO$_3$) and irradiated by a certain number of laser pulses $N$ per surface area. The peak laser fluence $F$ was chosen to be below the single-pulse ablation threshold fluence of Si in distilled water (F$_{th}\approx$ 0.11 J/cm$^2$) \cite{borodaenko2021deep}. Such fluence allows to avoid various uncontrollable effects deteriorating the laser surface nanopatterning as plasma formation, excessive bubbling and filamentation of the laser beam. Consequently, the single-pulse ($N$=1) irradiation caused only shallow ablation-free surface modification, where darker laser-exposed area in the SEM image appears to contain amorphous Si nanolayer (Figure 1b, left) \cite{garcia2016ultrafast,huang2019fabrication,huang2019cylindrically}. Within this area, NPs (averaged size of 15-40 nm) can be visually identified with their bright contrast associated with high electron emission inherent to metals. EDX also confirmed formation of Ag NPs within the laser-exposed area $via$ laser-induced reduction of Ag$^+$-cations to neutral metallic phase. The NPs are almost absent outside the exposed area highlighting that laser radiation localizes the redox reaction permitting targeted surface decoration. By increasing the number of applied pulse per surface spot, one can observe formation of Ag nanoparticle-embedded LIPSSs with their characteristic periods of $\approx$ 250 nm (at $N>$5) and $\approx$70 nm (at $N>$1000) (Figure 1b). Both types of LIPSSs follow nanotrench orientation perpendicular to the polarization vector of the incident laser radiation that is common for interference processes associated with surface waves.

Metallic NP formation under action of intense laser irradiation is commonly attributed to the generation of solvated electrons and hydrogen radicals produced during optical breakdown of water-based systems \cite{okamoto2019synthesis,ran2019femtosecond,broadhead2020fabrication,batista2021laser} or another active radical species depending on the composition of the solution \cite{herbani2011synthesis,meader2018radical}. However many precursors used for synthesis of NPs have the ability to be directly decomposed by laser even in relatively mild conditions far from the plasma formation as a result of photo- \cite{belmouaddine2017dense,mamonova2021single,jara2021photochemical} or thermo-induced processes \cite{smikhovskaia2019situ}. Here it can be reasonably assumed, that realization of both mechanisms is possible and contributes to the formation of NPs considering below-threshold pulse energies used for surface irradiation. The direct thermal decomposition similar to solid-state state reaction

\begin{equation}
2AgNO_3 = 2Ag + 2NO_2 + O_2
\end{equation}

\noindent is likely to be obstructed due to salt dissociation and formation of tetrahedrally coordinated complexes with water \cite{fox2002coordination}. This scenario can be considered in case of local heating within a focal spot up to the temperatures above 200 $^o$C for solvent evaporation and initiation salt decomposition according Eq. (1). Taking into account laser-induced melting of the silicon at 1410 $^o$C, the described processes can take place in the system under investigation. On the other hand, multiphoton absorption may cause the excitation of silver ions, which leads to transfer of electrons from the solvent molecules to Ag$^+$ with formation of neutral atoms (Eq. (2)) following by their agglomeration (Eq. (3)):

\begin{equation}
Ag^+ + H_2O \rightarrow Ag^0 + H^+ + \cdot OH
\end{equation}

\begin{equation}
nAg^0 \rightarrow (Ag^0)n
\end{equation}

Noteworthy, grating-type surface morphology allows to localize incident below-threshold laser energy at the bottom of the formed nanotrenches in the case when the polarization vector of the normally incident wave is oriented perpendicular to the LIPSSs (that is the case for our system; see Supporting Information). These electromagnetic ``hot spots'' allow to confine hardly distinguishable photo- and thermal-induced redox processes within nanoscale volume, localizing synthesis of the nanoparticles and particularly explaining their absence outside the nanotextured Si surface (Figure 1b).

\begin{figure}[t!]
\centering
\includegraphics[width=1.\columnwidth]{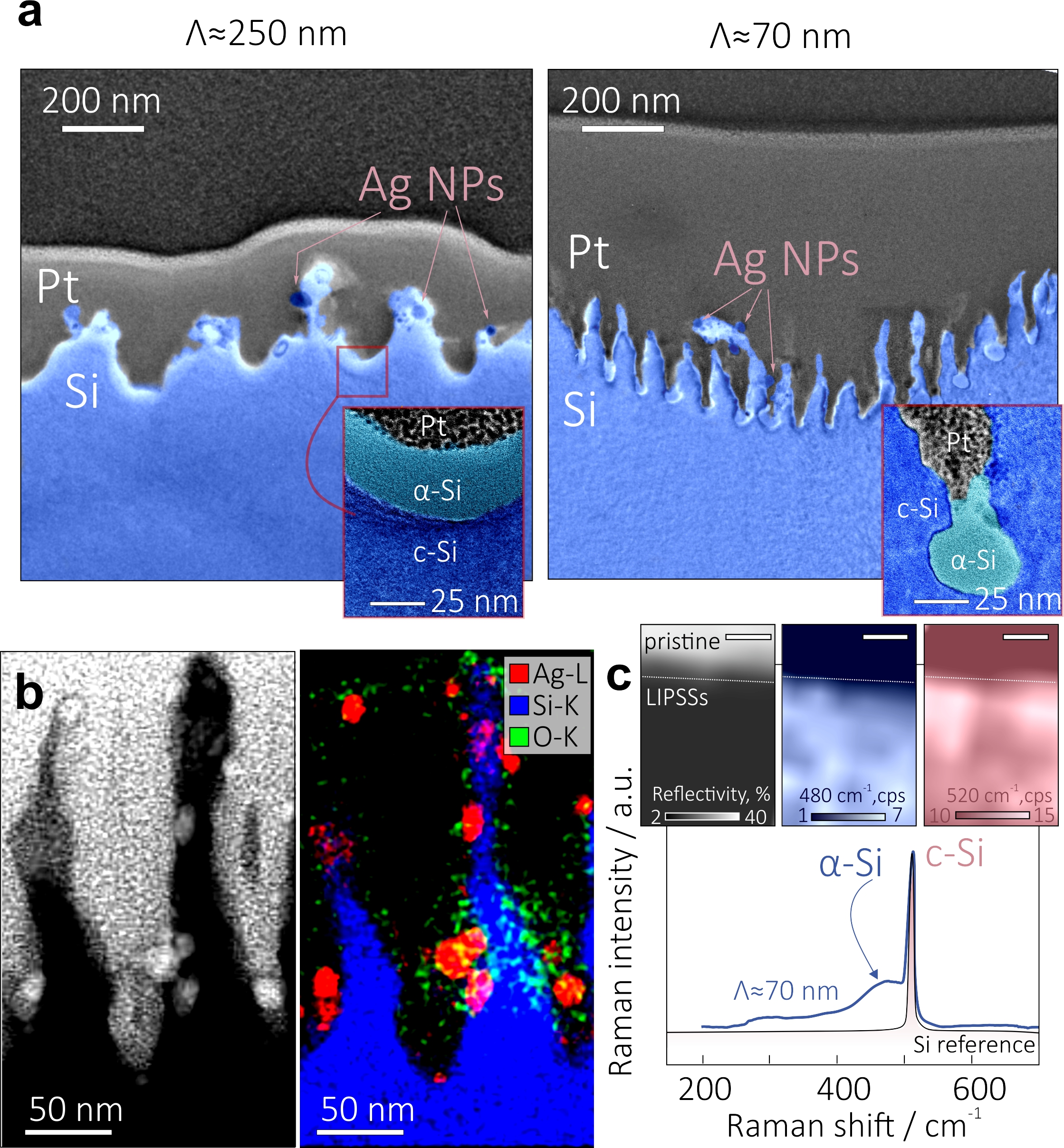}
\caption{(a) SEM images of cross-sectional FIB cuts revealing morphology of Ag-decorated LIPSS with two characteristic periods $\Lambda$=250 (left) and 70 nm (right). Bottom insets: HR-TEM images with well-seen amorphous Si shell in-between the adjacent nanospikes for both LIPSS types. (b) Correlated TEM image and EDX chemical composition map of the Ag-decorated LIPSSs with $\Lambda$=70 nm. (c) Representative Raman spectra of Ag-decorated LIPSSs ($\Lambda$=70 nm) and monocrystalline Si reference. Insets show reflectivity (at 473 nm pump) and micro-Raman maps (bands centered at 480 and 520 cm$^{-1}$) near the border between LIPSS area and pristine Si surface. Scale bar is identical for all maps and indicates 1 $\mu$m.}
\label{fig:2}
\end{figure}

Classic Sipe and van Driel model can be applied to explain the observed LIPSSs periods $\Lambda\approx 250-300$ nm appearing at $N$=5 - 25 \cite{sipe1983laser} . However, in a pure silicon-water system it predicts LIPSS periodicity $\Lambda\lesssim$200 nm, which is considerably shorter than that observed in our experiments, indicating that a thin modified layer must have appeared at the silicon-liquid interface. This layer can be assigned either to the photoexcited Si with the optical constants modified by increasing concentration of free electrons n$_e$ \cite{huangASCnano2009,derrien2014plasmonic}, or to the amorphization of the exposed surface \cite{Izawa2007APL}, or to the formation of metallic NPs (see inset in Figure 1c). The role of these three possible mechanisms for the modified layer formation will be analysed in the context of our experiments in the following tree paragraphs.

\begin{figure*}[t!]
\centering
\includegraphics[width=1.5\columnwidth]{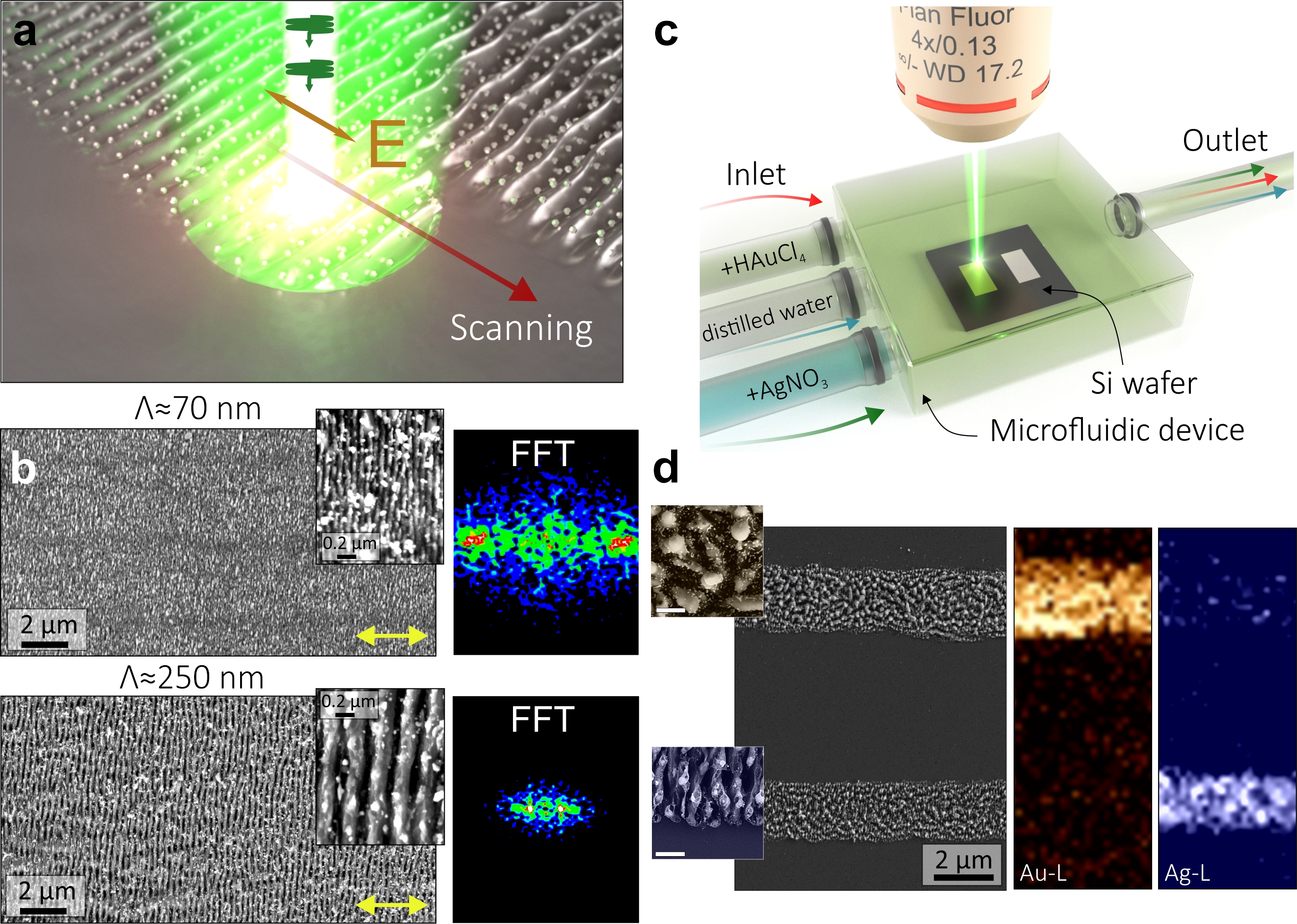}
\caption{
(a) Schematically illustrated fabrication of nanoparticle-decorated Si LIPSS area upon laser patterning of the sample in the functionalizing solution.
(b) Top-view SEM images of the Ag-decorated LIPSS produced by varying the scanning speed $v$: $v$=1  µm/s ($\Lambda$=70 nm) and 50 µm/s ($\Lambda$=250 nm). Insets provide close-up SEM and FFT images of the corresponding LIPSSs.
(c) Fabrication of the LIPSSs locally decorated by dissimilar NPs inside the microfluidic device allowing to refill the chamber with functionalizing solutions containing either HAuCl$_4$ or AgNO$_3$.
(d) SEM image of the neighboring LIPSS tracks decorated by Au and Ag NPs. The LIPSS were recorded by scanning the Si surface along line trajectory at $v$=50 $\mu$m/s, $F$=0.1 J/cm$^2$ and pulse repetition rate of 1 KHz. Functionalizing solutions containing either HAuCl$_4$ (5$\times$10$^{-5}$ M) or AgNO$_3$ (10$^{-3}$ M) were consequently ejected into the working chamber to produce each line. Insets illustrate nanoscale morphology of nanoparticle-decorated LIPSSs, while two right columns provide EDX maps of the same area showing distribution of Au and Ag elements.}
\label{fig:3}
\end{figure*}

Free charge carriers with concentration $n_e$ generated in the silicon due to photo-ionisation change the real and imaginary parts of the dielectric permittivity of crystalline silicon as $\varepsilon_r\approx 17.8-\dfrac{n_ee^2}{\varepsilon_om^*\omega^2(1+\nu^2/\omega^2)}$ and $\varepsilon_i\approx 0.06+\dfrac{n_ee^2\nu}{\varepsilon_om^*\omega^3(1+\nu^2/\omega^2)}$. Here $e$ is the electron charge, $m^*=0.16\times 10^{-30}$ kg is the optical mass of the electrons, $\nu\approx10^{15}$ Hz if the carrier collision frequency in Si \cite{Derrien2014rippled}, $\omega=2\pi c/\lambda$ is the incident light frequency. The real $n$ and imaginary $k$ parts of the refractive index change depending on the electron concentration as shown in Figure~\ref{fig:1}c. Depending of the irradiation conditions ($F$ and $N$) the experimentally observed LIPSS periodicity varies between 250 and 300 nm. This range of values can be explained by this photoexcited Si layer (Si$^*$) with the electron concentration n$_e$ varying from 1.8$\times$10$^{-28}$ to 3$\times$10$^{-28}$ m$^{-3}$ according to our estimations, see Fig.~\ref{fig:1}c.

In \cite{Izawa2007APL} a thin ($\sim10^{-8}$ m) amorphous silicon layer was observed on the laser-faced side of silicon after femtosecond laser processing. This layer appears due to the shock wave induced by laser irradiation, which amplitude exceeds 10-15 GPa \cite{DaisenbergerPRB2007polyamorphismSi}. As mentioned by \cite{miyaji_mechanism_2012}, the main role of the amorphous Si layer is to increase the absorption and consequently to facilitate the proto-excitation, however the optical constants of the amorphous silicon at 515 nm ($\varepsilon_r=18.94$, $\varepsilon_i=8.75$, \cite{AspensPRB1983AmorphSi}) also differ from those for the crystalline Si. Analogous to Figure~\ref{fig:1}c, the LIPSSs period can be calculated at the Si$^*$-Si interface \cite{derrien2014plasmonic}, the results do not differ much from that of Water-Si$^*$ system.

The role of the metallic NPs becomes more clear after comparing these results to our previous study \cite{borodaenko2021deep}. The NPs partly covering smooth Si surface are clearly seen in the Figure~\ref{fig:1}b (left). They form another surface layer with optical properties between those of the metal and the liquid on the top of the silicon wafer, the coverage depends on the efficiency of the redox reaction and hence, on the laser processing parameters. However, the same period $\Lambda\approx250$ nm was observed in silicon in \cite{borodaenko2021deep} with the same incident wavelength but without any NPs. This observation allows to conclude that the role of the pre-deposited NPs in the LIPSS period formation is negligible.
Cascaded reduction of the $\Lambda\approx 250-300$ nm LIPSSs period via ablative Rayleigh-Taylor instability developing in the liquid Si layer was recently justified to cause formation of the deep-subwavelength surface morphology with $\Lambda$=70 nm and orientation perpendicular to the polarization vector \cite{borodaenko2021deep}.

TEM imaging of the cross-sectional FIB cuts made perpendicularly with respect to the LIPSS nanotrenches being combined with EDX chemical mapping provides a deeper insight into structural features of the laser-textured surface. In particular, along with rather high aspect (depth-to-period) ratio of the protrusions ($\approx$ 1 and 3 for $\Lambda$=250 and 70 nm, respectively), the nanometer-thick shell containing disordered Si atoms can be observed near each surface elevation (Figure 2a). For both considered LIPSSs, the amorphous shell thickness is larger in between the elevations (bottom of the nanotrenches) as it is revealed by close-up inset TEM images (Figure 2a). This confirms the larger deposited laser intensity in this area owing to efficient localization of the EM field in the case when the polarization vector of the incident laser pulse is directed perpendicular with respect to the LIPSS trenches (see Supporting Information). Raman micro-spectroscopy confirmed the amorphous structure of the Si shell (characteristic $\alpha$-Si band at 480 cm$^{-1}$) at the bottom of the nanotrenches as well as polycrystalline structure of the LIPSS spikes (broader c-Si Raman peak at 520.8 cm$^{-1}$ as compared to the reference monocrystalline Si wafer, Figure 2c). Side walls of the nanospikes with a thinner disordered shell can also contain a certain amount of SiO$_2$ according to EDX mapping. Raman characterization also confirmed absence of residual AgNO$_3$ in the as-fabricated LIPSSs.

\begin{figure}[t!]
\centering
\includegraphics[width=1.\columnwidth]{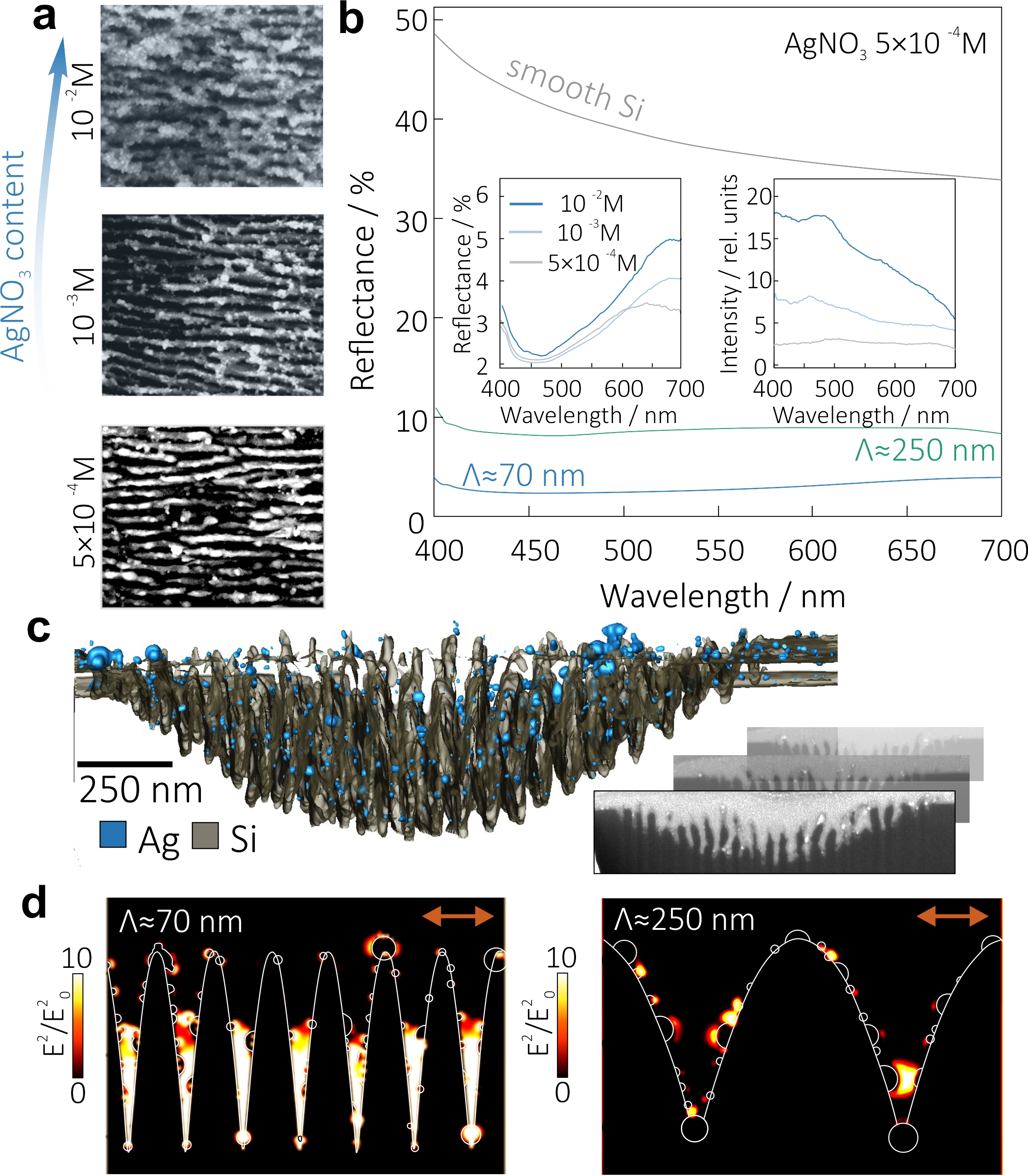}
\caption{
(a) Close-up SEM images of the Ag-decorated LIPSSs ($\Lambda$=70 nm) produced with 5$\times$10$^{-4}$, $10^{-3}$ and $10^{-2}$ M aqueous solutions of AgNO$_3$.
(b) Reflectance spectra of the Ag-decorated LIPSSs ($\Lambda$=70 nm and $\Lambda$=250 nm produced with 5$\times$10$^{-4}$M AgNO$_3$ aqueous solutions) and smooth Si reference. Insets show polarized reflectance and back-scattered spectra of the Ag-decorated LIPSSs ($\Lambda$=70 nm) produced with 5$\times$10$^{-4}$, $10^{-3}$ and $10^{-2}$ M aqueous solutions of AgNO$_3$. Inset reflectance spectra were obtained under excitation with white light polarized perpendicular with respect to LIPSSs orientation.
(c) 3D model of the Ag-decorated Si LIPSSs ($\Lambda$=70 nm) reconstructed from a series of SEM images of the cross-sectional FIB cuts.
(d) Normalized squared EM field amplitude ($E^2$/$E_0^2$) calculated near Ag-decorated Si LIPSSs ($\Lambda$=70 nm and $\Lambda$=250 nm) irradiated from the top by x-polarized plane wave at 473 nm.}
\label{fig:4}
\end{figure}

The laser-printing approach allows to replicate the desired type of the LIPSSs over a large surface area by scanning the Si wafer placed into functionalizing solution with a laser beam at constant speed $v$  (Figure 3a). Both types of the nanoparticle-embedded surface morphologies with $\Lambda$=250 and 70 nm obtained at fixed laser fluence $F$=0.1 J/cm$^2$ by varying $v$ are provided in Figure 3b. As can be seen, along with reduction of the characteristic LIPSSs period $\Lambda$, decrease of the scanning speed results in larger number of incident pulses per surface area thus increasing the amount of the generated Ag NPs. EDX studies confirmed increasing amount of Ag by analyzing and averaging signal collected from the identical LIPSS surface areas with the size of 100 $\times$100 $\mu$m$^2$. The performed experiments also clearly show that the formation of the decorating nanoparticles via laser-induced reduction can be efficiently localized within the laser focal spot. This opens up prospects for fabrication of the nanotextured Si surface areas locally decorated with dissimilar plasmon-active materials. To illustrate this remarkable modality, the experiments were carried out in home-built microfluidic device allowing to replace functionalizing solutions for laser texturing as well as clean the working chamber by pure distilled water (Figure 3c). To illustrate the key idea of the approach we used aqueous solutions containing either AgNO$_3$ (10$^{-3}$ M) or HAuCl$_4$ (10$^{-5}$M). First, owing to galvanic replacement surface texturing was performed to create LIPSSs decorated by Au NPs in the HAuCl$_4$ solution. Next, the working chamber with the sample was cleaned by pure distilled water several times following by refilling with the second functionalizing solution containing AgNO$_3$ (10$^{-3}$ M). SEM imaging combined with EDX mapping confirmed formation of the LIPSSs locally decorated by dissimilar plasmonic NPs (Figure 3d). Noteworthy, small amount of Ag species was detected within the Au-decorated LIPSSs. This indicates that the Ag NPs were synthesized in the focal volume irradiating the functionalizing solution and preferentially deposited within the LIPSS areas owing to their strong wetting behaviour. Non-textured Si surface areas are almost free of metallic NPs after the cleaning with distilled water. The approach allows to create other practically relevant metal-semiconductor compositions. The similar experiments were carried out with functionalizing solutions containing H$_2$PdCl$_4$ to produce Si LIPSSs decorated with Pd NPs (see Supporting Information).

\subsection{Optical and sensing properties nanoparticle-embedded Si LIPSSs}\label{Sec:measurements}

\begin{figure}[b!]
\centering
\includegraphics[width=1.\columnwidth]{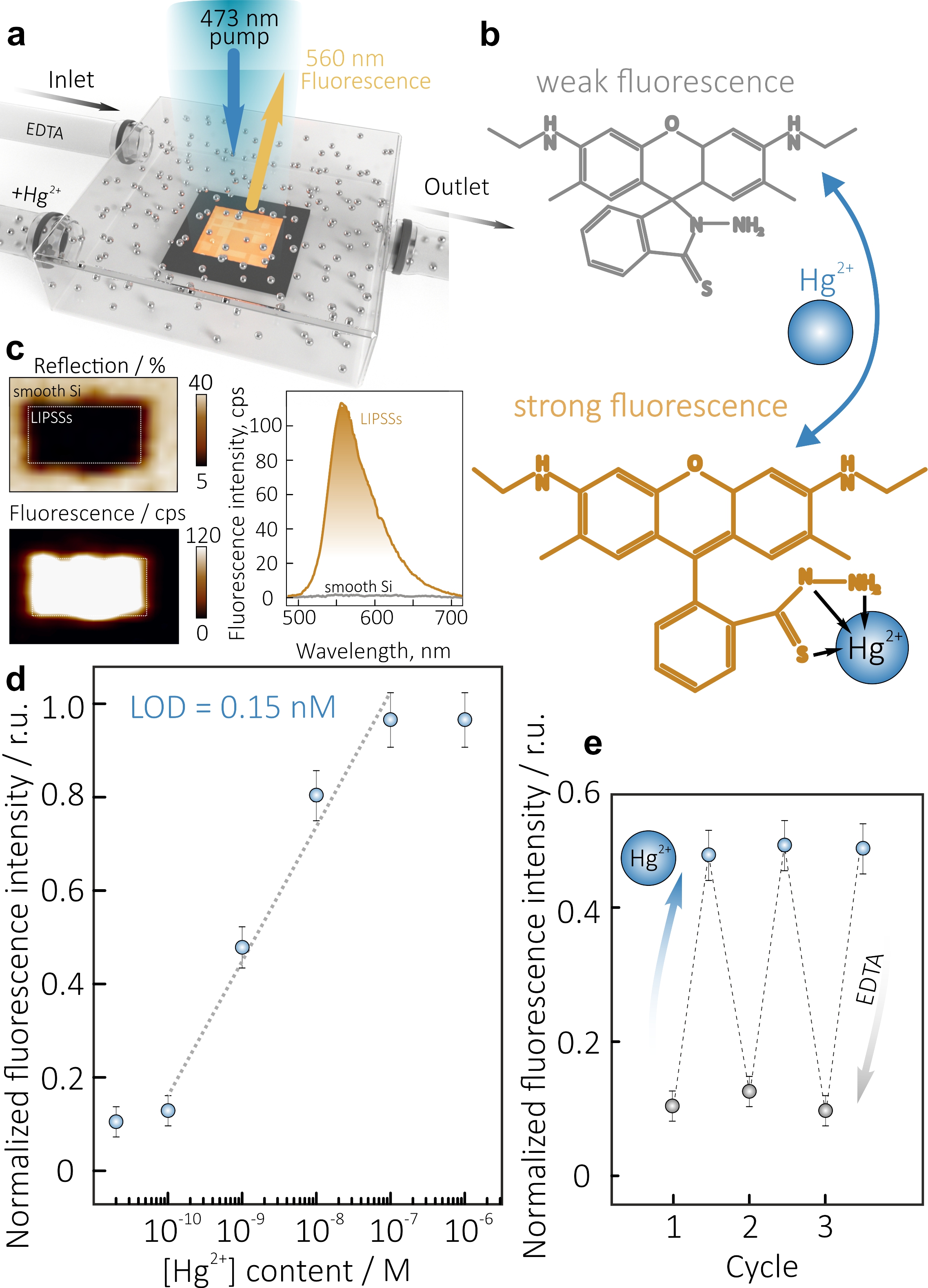}
\caption{(a) Schematic of the experimental setup used for Hg$^{2+}$ optical nanosensing and (b) structure of the Rhodamine 6G thiohydrazide revealing the origin of the fluorescence response. (c) Correlated fluorescence (560$\pm$30 nm) and confocal reflection (473 nm) maps of the deep-subwavelength LIPSS area functionalized with Rhodamine 6G thiohydrazide molecular nanolayer. Fluorescence spectra highlight difference of the optical signal measured on laser-textured and smooth Si areas. (d) Normalized fluorescence intensity versus the Hg$^{2+}$ concentration in the distilled water. (e) Reversibility of the sensor's optical response upon several cycles of analyte injection and washing.
}
\label{fig:5}
\end{figure}

Laser processing parameters allow to control the desired type of the Si surface morphology. Moreover, by tuning the amount of the AgNO$_3$ in the functionalizing solution one can tailored the corresponding density (amount) of the decorating plasmon-active NPs per surface area. In fact, the method can be considered as a template-assisted chemical synthesis, where high-aspect-ratio LIPSS nano-trenches serve as cages for plasmonic NPs. Figure 4a illustrates this useful modality by providing close-up SEM images of Ag-decorated Si LIPSSs ($\Lambda$=70 nm) produced by increasing the amount of AgNO$_3$ from 5$\times$10$^{-4}$ to $10^{-2}$ M. Noteworthy, LIPSS patterning of the Si wafer typically leads to reduction of the reflectivity within broad spectral range as a result of decrease refractive index jump at the air-Si interface \cite{shen2008high}. LIPSSs with both observed periods allow to decrease reflectivity (with respect to those for pristine Si wafer). In particular, when probed with unpolarized white light, the averaged visible-range reflectivity drops to $\approx$ 3 and 9\% for LIPSSs with $\Lambda$=70 and 250 nm (Figure 4b). Interestingly, when the white light is polarized perpendicularly to the nanotrenches, local drop of the reflectivity can be observed for the Ag-decorated LIPSSs indicating that the most efficient coupling of the incident waves to localized EM hot spot can be expected for blue-green spectral range (left inset, Figure 4b). In this spectral range, plasmon-active Ag NPs decorating the nanotrench walls can be efficiently pumped further boosting the intensity of the EM hot spots within the air gaps. Correlated dark-field (DF) back-scattering spectra measured from the Ag-decorated deep-subwavelength LIPSSs indicate the broad scattering maximum centered at 480 nm that can be attributed to excitation of localized plasmon resonance of the embedded Ag NPs (right inset, Figure 4b). Substrate effect and plasmon mode hybridization reasonably explains strong red-shift of the localized plasmon resonance of the Ag NPs efficiently interacting with the high-refractive-index Si. Noteworthy, elevating amount of the decorating NPs achieved by tuning the AgNO$_3$ content in the functionalizing solution increases both the reflectivity in the red spectral band as well as intensity of the characteristic plasmon-mediated scattering (inset, Figure 4b).

Altogether, low surface reflectivity and plasmonic activity of the embedded NPs provide optimal conditions for localization and enhancement of the EM fields within the air gaps between metal-decorated Si protrusions. This deduction was confirmed by finite-difference time-domain (FDTD) simulations of the squared EM field amplitude ($E^2$/$E_0^2$) calculated near Ag-decorated LIPSSs with $\Lambda$=70 and 250 nm upon their excitation from the top by a plane wave at 473 nm pump (that is close to the localized plasmon resonance of the Ag NPs on the Si protrusions revealed by DF spectroscopy). For more reliable modeling, 3D models of the Ag-decorated LIPSSs were reconstructed from the series of SEM images of the cross-sectional FIB cuts. Example of such model reconstructed for LIPSSs with $\Lambda$=70 nm where the Si and Ag materials are highlighted by different colors is given in Figure 4c. The performed simulations revealed substantial (at least, tenfold) enhancement of the $E^2$/$E_0^2$ value can be achieved for both characteristic morphologies in the case when the polarization vector of the pump radiation is oriented perpendicularly with respect to the LIPSS orientation. Meanwhile, deep-subwavelength LIPSSs provide extremely high density of such hot EM spots is a sharp contrast to surface morphology with $\Lambda$=250 nm. Noteworthy, for surface morphology with $\Lambda$=70 nm the characteristic $E^2$/$E_0^2$ map was rescaled for better comparison with those for $\Lambda$=250 nm, while the maximal EM field enhancement in the certain spots reaches 10$^2$ highlighting applicability of such LIPSSs for optical nanosensing. Stronger loading of the Si nanotrenches allows to occasionally produce smaller air gaps exhibiting substantial field enhancement (Supporting Information).

Mentioned optical properties of Ag-decorated deep-subwavelength Si LIPSS make this easy-to-fabricate surface morphology promising for various applications utilizing  interaction of localized EM hot spots with various quantum emitters (organic molecules, quantum dots, etc.). Previous reports highlight applicability of the metal-decorated Si LIPSSs as advanced substrates for SERS-based molecular nanosensing \cite{hamad2018femtosecond,li2020shaped,borodaenko2021deep}. Here, we expand the application range of the metal-semiconductor nanotextures by demonstrating their applicability for optical sensing based on surface enhanced fluorescence (SEF) effect \cite{geddes2002metal}. To do this, the laser-textured Si wafers containing Ag-decorated LIPSSs ($\Lambda$=70 nm) were first functionalized with a Hg$^{2+}$-sensitive fluorophore (Figure 5a), Rhodamine 6G thiohydrazide (further referred to as d108), rendering the substrate selective ionochromism (Figure 5b). Arising from the efficient coupling of the normally incident linearly polarized radiation to the air gaps containing Ag nanoparticles and d108 emitting molecules, we observed stable and detectable fluorescence signal of the functionalized LIPSSs even at their low-intensity ($\approx$ 1 $\mu$W/$\mu$m$^2$) laser pump at 473 nm wavelength that ideally fits both the plasmon band of the NPs and absorption band of the emitters.

Figure 5c provides correlating confocal reflectivity (at 473 nm laser pump wavelength) and fluorescent (at 560$\pm$30 nm fluorescence wavelength) maps clearly demonstrating uniformity of the signal enhancement over the 100 x 50 $\mu$m$^2$ surface area patterned by deep-subwavelength Ag-decorated Si LIPSSs. Such LIPSSs were found to demonstrate at least 55-fold enhanced fluorescence signal (depending of the amount of the Ag NPs) with respect to those observed on surrounding smooth Si surface areas. Moreover, low-intense excitation provided negligible ($\approx$ 3\%) degradation of the fluorescence intensity upon making multiple signal maps. Such remarkable photostability as well as low water solubility of the d108 ensured reliable fixation and non-leaching of the sensor layer during multiply injection of the distilled water containing trace amount of Hg$^{2+}$ ions (Figure 5a). In its turn, the LIPSS-based sensor reacts by increasing the average fluorescence intensity plotted on the Figure 5d as a function of mercury ion concentration in the injected liquid. Systematic studies of the device response allowed to identify the standard limit of detection ($\approx$0.15 nM) as well as the range of Hg$^{2+}$ ion concentrations providing linear sensor response (in log scale) from 0.1 to 100 nM. The mercury ions can be successfully removed from sensitive layer by sequential washing of the sample in the EDTA solution and distilled water, as a result the sample restored its initial signal level demonstrating reusability and reversibility of the optical response (Figure 5e).

In addition to the obvious advantage of reusability, these results clear illustrate the performance benefits of the SEF-based sensing approach. The sensing performance of soluble dye-based probes depends on their brightness (absorptivity $\times$ fluorescence quantum yield), as well as the equilibrium constant of probe-analyte interaction. Limited by molar absorptivity $\approx 10^5$ and binding constant $\approx 10^3 - 10^5$, real probes show typical working range of sensor operation of $\approx$ 0.1 $\mu$ M - 0.1 mM, which is much higher than the maximum permissible concentration of the most toxic metal ions like Hg$^{2+}$. Although the used probe (like many sulfur-containing compounds) has extremely high affinity toward mercury ions, the reliable determination of analyte at low concentrations is hardly possible due to insufficient brightness of the ``analyte-probe'' complex and the inability to accumulate sufficient signal from extremely dilute solution. Thus, ``standard'' measurements in solution revealed the limit of detection $\approx$ 0.05 $\mu$M, while optical measurements from a densely packed and surface-enhanced sensing layer demonstrated 300-fold better performance.

\section{Conclusions}

Here, direct multi-pulse femtosecond laser processing of crystalline Si wafer in distilled water containing AgNO$_3$ and HAuCl$_4$ at millimolar concentrations was demonstrated for fabrication of LIPSSs decorated by plasmon-active NPs. LIPSSs on a Si surface with a variable periodicity ranging from 300 to 70 nm and controllable amount of embedded plasmon-active NPs were produced and comprehensively characterized by SEM, TEM, EDX analysis, optical and Raman spectroscopy. Hybrid nano-morphologies representing 70-nm period nanogratings on the Si surface with their 200-nm depth nanotrenches loaded with plasmonic NPs allow efficient light localization in the air gaps providing pathways for optical nano-sensing applications. As a demonstration of this statement, we realized SEF-based fluorescence detection of trace concentrations (down to 0.15 nM) of Hg$^{2+}$ ions in water.

Formation of plasmonic NPs was found to be localized within the laser focal spot through the laser-induced chemical reduction allowing to produce LIPSS locally decorated with dissimilar noble-metal nanoparticles. This remarkable demonstration highlights the capability of the proposed fabrication approach to produce advanced ``pixelaed'' optical sensors (or sensor micro-arrays), where each ``pixel'' can be functionalized with various metal species to cover multiple analytes and excitation wavelengths.

\section{Methods}

\subsection{Materials and chemicals}
p-type boron-doped Si wafers (0.5$\times$1 cm$^2$) with specific resistivity of 10-20 ?$\cdot$cm$^{-1}$, (111) oriented surface and 0.5 mm thickness were used as a substrates for laser nanotexturing. The substrates were sonicated in NH$_3$/H$_2$O$_2$/H$_2$O (1/1/3, v/v/v) solution at 80$^o$C for 30 minutes, thoroughly rinsed with deionized water and dried at 100$^o$C before use.

Silver nitrate (Sigma-Aldrich, 99\%), Gold(III) chloride trihydrate (Sigma-Aldrich, 99.9\%), Mercury(II) nitrate monohydrate (Sigma-Aldrich, 98.5\%), Rhodamine 6G (Sigma-Aldrich, 99\%), Lawesson's reagent (Sigma-Aldrich, 97\%), hexane (Sigma-Aldrich, 95\%), chloroform (Sigma-Aldrich, 99\%), ethyl acetate (Sigma-Aldrich, 99.8\%), hydrazine hydrate (Sigma-Aldrich, 98\%), silica gel (100/200 ?m) were used as received. NMR spectra were recorded on the Bruker Avance 400 with the frequency of proton resonance 400 MHz using CDCl$_3$ as the solvent.

\subsection{Fluorescent probe synthesis}

Rhodamine 6G hydrazide was prepared as described elsewhere \cite{zhang2010naphthalimide}. Yield 80\%. $1$H NMR (400 MHz, CDCl$_3$, ppm, ?): 7.96 (m, 1H), 7.45 (m, 2H), 7.06 (m, 1H), 6.39 (s, 2H), 6.26 (s, 2H), 3.58 (s, 2H), 3.54 (br.s, 2H), 3.22 (q, 4H), 1.92 (s, 6H), 1.32 (t, 6H); Elemental Analysis data: Calc. C, 72.87; H, 6.59; N, 13.07; Expt. C, 72.97; H, 6.66; N, 12.89.

Rhodamine 6G thiohydrazide (d108) was synthesized by the procedure similar to Rhodamine B thiohydrazide \cite{zheng2006switching}. Rhodamine 6G hydrazide (60mg, 0.14 mmol) and Lawesson's reagent (57mg, 0.14 mmol) were dissolved in dry toluene. The mixture was degassed, purge with argon and refluxed with stirring for 6 h. After cooling to room temperature, the solvent was removed at 80$^o$C under reduced pressure. A crude product was purified by flash column chromatography on a silica gel with hexane/ethyl acetate (v/v = 1/1) as the eluent to afford the product as a pink powder (42.3 mg, 68\%). $^1$H NMR (400 MHz, CDCl$_3$, ppm, ?): 7.98 (m, 1H), 7.42 (m, 2H), 7.06 (m, 1H), 6.38 (s, 2H), 6.24 (s, 2H), 3.61 (s, 2H), 3.58 (br.s, 2H), 3.20 (q, 4H), 1.92 (s, 6H), 1.32 (t, 6H); Elemental Analysis data: Calc. C, 70.24; H, 6.35; N, 12.60; Expt. C, 70.29; H, 6.36; N, 12.71.

\subsection{Laser nanotexturing}
Laser nanotexturing of a monocrystalline Si wafer placed in a quartz cuvette was carried out by 200-fs second-harmonic (513-nm wavelength) linearly-polarized laser pulses generated by regeneratively amplified Yb:KGW laser system (Pharos, Light Conversion). Laser pulses were focused onto the Si surface through a few-mm thick liquid layer (distilled water containing a certain variable amount of HAuCl$_4$ or AgNO$_3$) using a long-focal-distance objective with a numerical aperture of 0.13 providing a 1/e$^2$-diameter of the focal spot $\approx$ 5 $\mu$m. Fixed pulse repetition rate of 1 KHz was used in all experiments to avoid excessive liquid bubbling. The cuvette was placed onto a motorized nanopositioning system (ANT Series, Aerotech) to pattern the Si surface areas along the computer-defined trajectory. Pyroelectric detector (Ophir) was used to assess the laser fluence incident onto the Si surface.

\subsection{Structural Characterization}
Nanoscale morphology and composition of the nanoparticle-decorated Si LIPSS were analyzed using SEM (Carl Zeiss Ultra 55+) and TEM (Titan 60-300, Thermo Fisher). Both facilities were equipped with EDX detectors for chemical composition characterization. For high-resolution TEM studies, a 100-nm thick lamellas were prepared using FIB milling procedure (Helios 450, Thermo Fisher) that includes a coverage of the region of interest by a protective Pt over-layer. Also, a series of consecutive FIB slices (each visualized by SEM) was prepared $via$ FEI AutoSlice View G3 software  to perform the 3D morphology reconstruction. The obtained slices were further processed with Avizo 8.1 software (Thermofisher) to create the 3D model of the LIPSSs where the Ag NPs and Si nanotextures can be visually distinguished.

\subsection{Optical characterization and sensing}

A home-built optical setup confocally connecting the optical microscope (Eclipse Ti, Nikon) and a grating-type spectrometer (Shamrock 303i, Andor Technologies) equipped with a thermoelectrically cooled CCD-camera (Newton, Andor Technologies) was used to measure the polarization-resolved reflectivity and dark-field back-scattering spectra of nanoparticle-decorated LIPSSs. Both measurements were carried out with a halogen lamp and a dry optical objective collecting reflected/scattered signal (NA=0.8, 100x Mitutoyo).

Micro-Raman spectroscopy setup (Spectra II, NT-MDT) was implemented to study crystalline structure of the LIPSSs as well as to assess their applicability for fluorescent-based biosensing. In both cases, pump laser radiation (473 nm wavelength) was focused onto the sample surface with a dry lens (NA=0.7, 100x Mitutoyo). The spectra were analyzed by a spectrometer equipped with a thermoelectrically cooled CCD-camera (i-Dus, Andor Technologies). Mapping of Raman/fluorescence signal distributions over the laser-textured (smooth) surface sites was realized by a galvanic scanner. Pump laser intensity was adjusted by a built-in attenuator to avoid fluorescence degradation or excessive heating. To investigate SEF sensing potential Ag-decorated LIPSSs were functionalized by immersing into d108 tetrahydrofuran solution (1 mM) for 10 min followed by rinsing with pure solvent and air drying. The microfluidic device was used to inject either Hg$^{2+}$ aqueous solution with variable content of ions or EDTA solution to refresh the sensor response.

Supporting FDTD simulation were carried out with commercial electromagnetic FDTD solver (Lumerical, Ansys). 2D distributions of the squared electromagnetic field amplitude (E$^2$) normalized onto the amplitude of the incident field (E$_0^2$) were simulated under excitation of the LIPSS structure by normally incident linearly polarized plane wave at 473 nm and 515 nm pump wavelengths. The 3D surface morphology for modeling was reproduced from the corresponding reconstruction of the  was reproduced from corresponding SEM images of cross-sectional cuts. Computational volume with a 1x1x1 nm$^3$ mesh was limited by perfectly matched layers.

\section*{Conflicts of interest}
The authors declare no conflict of interest.

\section*{Data Availability Statement}
The data that support the findings of this study are available from the
corresponding author upon reasonable request.

\section*{Keywords}
Femtosecond laser pulses; Laser-induced periodic surface structures; Metal-semiconductor nanostructures;  Optical sensing; Surface-enhanced fluorescence.

\section*{Acknowledgements}
This work was supported by the Russian Science Foundation (Grant no. 21-79-20075).

\end{document}